\documentclass{PoS}

\usepackage[utf8]{inputenc}
\usepackage[OT2,T1]{fontenc}
\usepackage{amssymb,times,mathptmx,graphicx}
\usepackage{bm,bbm}

\newcommand{\be}{\begin{equation}}
\newcommand{\ee}{\end{equation}}
\newcommand{\ba}{\begin{eqnarray}}
\newcommand{\ea}{\end{eqnarray}}
\def\bs{\begin{subequations}}
\def\es{\end{subequations}}
\def\a{\alpha}

\def\g{\gamma}
\def\la{\lambda}

\def\om{\omega}

\def\cP{\mathcal{P}}

\def\cV{\mathcal{V}}

\def\ds{d_{\rm S}}
\def\dh{d_{\rm H}}

\newcommand{\Eq}[1]{(\ref{#1})}

\newcommand{\oarX}[1]{\href{http://arxiv.org/abs/#1}{{arXiv:#1}}}
\newcommand{\arX}[1]{\href{http://arxiv.org/abs/#1}{{arXiv:#1}}}

\newcommand{\doin}[6]{\href{http://dx.doi.org/#1}{{\it #2 #3} {\bf #4} (#6) #5}}
\newcommand{\doinn}[5]{\href{http://dx.doi.org/#1}{{\it #2} {\bf #3} (#5) #4}}
\newcommand{\doij}[5]{\href{http://dx.doi.org/#1}{{\it #2} {\bf #3} (#5) #4}}

\newcommand{\books}[5]{{\it #1}, #2, #3 #5}
\newcommand{\tia}[1]{\emph{#1},}

\def\lp{\ell_{\rm Pl}}

\def\rme{\textrm{e}}
\def\rmd{\textrm{d}}

\title{Detecting quantum gravity in the sky}

\ShortTitle{Detecting quantum gravity in the sky}

\author{\speaker{Gianluca Calcagni}\\
        Instituto de Estructura de la Materia, CSIC, Serrano 121, 28006 Madrid, Spain\\
        E-mail: \email{g.calcagni@csic.es}}

\abstract{Getting signatures of quantum gravity is one of the topical lines of research in modern theoretical physics and cosmology. This short review faces this challenge under a novel perspective. Instead of separating quantum-gravity effects of a specific model between UV and IR regimes, we consider a general feature, possibly common to many frameworks, where all scales are affected and spacetime geometry is characterized by a complex critical exponent. This leaves a log-oscillating modulation pattern in the cosmic microwave background spectrum and gives a unique opportunity, illustrated with the example of a multi-fractional theory, to test quantum gravities at cosmological scales.}

\FullConference{EPS-HEP 2017, European Physical Society conference on High Energy Physics\\
		5-12 July 2017\\
		Venice, Italy}

\begin{document}

Despite its one hundred years of success, general relativity has been questioned as a fundamental theory for the big-bang and black-hole singularities it is plagued by, as well as for its inability to explain satisfactorily the cosmological constant problem. There are many ways to go beyond Einstein gravity: some simply modify the classical dynamics of the metric, while others invoke quantum mechanics and attempt to quantize gravity hoping that the uncertainty principle could resolve the classical infinities and, why not, generate new phenomenology. Finding a consistent theory of quantum gravity is a challenge that has found several positive answers \cite{Ori09,Fousp}. However, probing such a theory with presently available experimental data turns out to be difficult because quantum effects are usually small, even in the early universe where the Planck scale is inflated to cosmic size. One identifies two ranges of scales, the infrared (IR) where classical standard physics is recovered and the ultraviolet (UV) where short (sub-Planckian?) scales are characterized by exotic model-dependent effects. Calling $\ell_*$ (typically, of order of the Planck scale $\lp$) the scale of reference separating the UV from the IR, these effects are suppressed by the ratio between $\ell_*$ and a much larger observation-dependent scale. Does the UV/IR divide dominate all possible effects that can arise when merging general relativity with quantum mechanics? We argue for a negative answer that bears consequences for the observability of new physics. 

Let us recall a general feature found in all quantum-gravity theories, namely, the fact that, once the notion of spacetime is built from the underlying fundamental (pre)geometric structure, the dimension of spacetime is well defined and changes with the probed scale (see, e.g., \cite{revmu,Car17}). One can have different definitions of dimension in mind. The Hausdorff dimension $\dh$ is the scaling of the volume $\cV$ of a ball or hypercube with respect to its radius or edge length $\ell$; when $\dh$ is constant, $\cV\sim\ell^{\dh}$. The spectral dimension $\ds$ is the scaling of the return probability $\cP$, a quantity derived from a diffusion process representing the probing of the geometry with an experimental resolution $1/\ell$; when $\ds$ is constant, $\cP\sim\ell^{-\ds}$. The great majority of quantum gravities experience a variation in the spectral dimension, while the Hausdorff dimension is unchanged.

Different quantum gravities have a different profile $\ds(\ell)=-\rmd\ln\cP(\ell)/\rmd\ln\ell$: the UV value may change, above or below the topological dimension $D$, while there may be intermediate plateaux, peaks or valleys. The surprising (at least for the author) thing is that, as soon as the spectral dimension is scale-dependent, there exists a universal parametric form of this profile. For reasons of space, here we will limit our attention to a configuration with no intermediate plateaux and with the same anomalous scaling of time and spatial directions. Then, the spectral dimension of a quantum gravity with these characteristics comes from the return probability \cite{first}
\be
\cP(\ell) =\frac{1}{\ell^D}\left[1+\left(\frac{\ell_*}{\ell}\right)^\a F_\om(\ell)\right]\,,\label{cP}
\ee
where the fractional exponent $\a$ is a real positive parameter, $F_\om$ is the log-oscillating modulation factor
\be
F_\om(\ell)=A_0+\sum_{n=1}^{+\infty}\left[A_n\cos\left(n\om\ln\frac{\ell}{\ell_\infty}\right)+B_n\sin\left(n\om\ln\frac{\ell}{\ell_\infty}\right)\right],
\ee
$A_n$ and $B_n$ are real amplitudes between 0 and 1, $\om$ is a frequency corresponding to the imaginary part of the spectral dimension (also called complex spectral dimension) \cite{cmplx}, and $\ell_\infty$ is another length scale that can be set to be equal either to $\ell_*$ or to the Planck length \cite{revmu}. Although there may be other intermediate scales below $\ell_*$, they do not affect the IR limit. In the UV, $\ds\simeq D\a$.

Thus, while inequivalent theories give rise to different parameters $\Pi=(\a,\om,A_n,B_n,\ell_*,\ell_\infty)$, all predict the same parametric form \Eq{cP} of the return probability. The first and foremost consequence of this result is that the scale hierarchy of geometry is not limited to $\ell_*$, the divide between UV and IR physics, but it spreads over an infinite sequence of scales $\la_\om^{m}\ell_\infty$, $m\in\mathbb{Z}$, where $\la_\om:=\exp(-2\pi/\om)$ is completely determined by the complex dimension $\om$. The origin of this sequence is a discrete scaling symmetry $\ell\to\la_\om\ell$ in the UV that leaves $F_\om$ invariant. In other words, a spacetime with complex spectral dimension is discrete in the UV and has a hierarchy of infinitely many scales, some of which are understood as ``large'' ($\la_\om^{m}\ell_\infty>\ell_*$ for a given $m$) and the others are ``small'' ($\la_\om^{m}\ell_\infty<\ell_*$). To summarize, some underlying model-dependent quantum-gravity mechanism generates a flow of the dimension $\ds(\ell)$ with the probed scale, which in turn produces (if the symmetries of the model allow it) a \emph{long-range} modulation of geometry that can be detectable at any scale, from the laboratory to the galaxies. If the parameters $\Pi$ predicted by the theory take fortunate values and, in particular, the amplitudes $A_n$ and $B_n$ do not vanish simultaneously, then there is a hope to observe quantum gravity where it was previously believed it would be impossible.

Of course, constraints that apply to general relativity and to the Standard Model of particles already limit the range of these parameters in such a way that any exotic effect is bound to be small. However, there is much room for new phenomenology, which we illustrate with the example of multi-fractional spacetimes \cite{revmu}. These are a framework where the position- and momentum-space measures implement dimensional flow via the universal parametrization discussed above, without fixing the value of the parameters until the very end. In this sense, multi-fractional theories are a sort of \emph{passe-partout} by which one can study certain regimes of quantum gravity in a model-independent way. However, due to the assumption of factorizability of the measures, these theories have features of their own not shared by other models, and they can also be regarded as stand alone. 

Depending on the symmetries of the Lagrangian, there are three different proposals. Here we focus on the theory with so-called $q$-derivatives, where comoving momentum space has a measure $\rmd^Dp(k)=\prod_\mu\rmd p^\mu(k^\mu)$ with $p^\mu(k^\mu)=k^\mu[1+\a^{-1}({k^\mu}/{k_*})^{1-\a}F_\om(k^\mu)]^{-1}$. The primordial power spectrum of scalar perturbations generated during an early stage of inflationary expansion takes the form $P_{\rm s}(k)\propto[p(k)]^{n_{\rm s}-1}$, where $p(k)=|{\bf p}|$ and $n_{\rm s}$ is the scalar spectral index. Spacetime discreteness at scales $\sim \ell_\infty$ may be potentially visible as a logarithmic modulation of the power spectrum of primordial fluctuations. This possibility has been tested with cosmic microwave background (CMB) data by considering only the first harmonic $n=1$ \cite{frc14}. Curiously, an $\om$-dependent upper bound $\a\lesssim 0.1-0.6$ was found for the fractional exponent, leading to an estimate of the Hausdorff and spectral dimension of space in the UV, valid as long as there is a non-trivial dimensional flow:
\be
\dh^{\rm \,space},\ds^{\rm \,space}=3\a\lesssim 0.3-1.9\qquad\textrm{(UV)}\,.
\ee
Upper bounds on $A_1$ and $B_1$ have also been found, $A_1,B_1<0.4$. CMB bounds on $\ell_*$ are weak and one should use other observations, in particular astrophysical, to place constraints on that scale \cite{revmu}. 

\begin{figure}[ht]
\centering
\includegraphics[width=13cm]{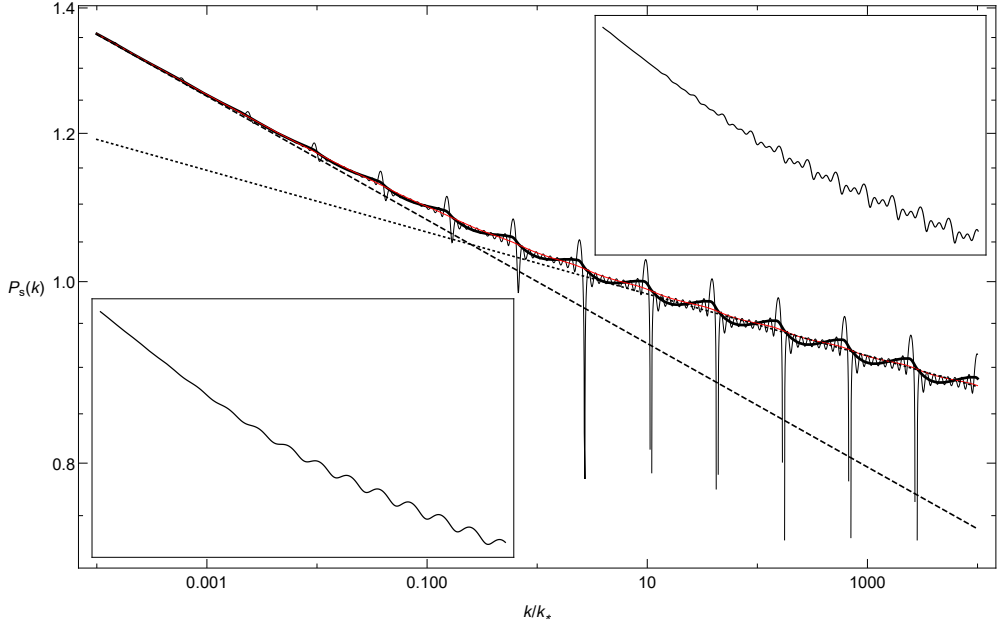}
\caption{\label{fig1} Main plot: log-log plot of the multi-fractional primordial scalar power spectrum for $n_{\rm max}=9$ and $\g=0=u$ (thin black curve), $\g=0$ and $u=2$ (thick black curve), and $\g=2$ and $u=0$ (red curve), compared with the standard spectrum (dashed line) and the deep-UV trend $k^{\a(n_{\rm s}-1)}$ (dotted line). Insets: case $\g=0=u$ with $n_{\rm max}=1$ (lower plot) and $n_{\rm max}=3$ (upper plot). The other parameters are $a_n=b_n=0.2$, $k_\infty=1/\lp\approx 10^{57}\textrm{Mpc}^{-1}$ \cite{revmu,frc14}, $k_*=1$, $\om=2\pi\a/\ln 2$ \cite{frc14}, $n_{\rm s}=0.967$ \cite{P1520}, $\a=1/2$. While the bend in the spectrum is a typical UV/IR effect, the oscillatory pattern is due to discrete scale invariance. Credit: \cite{cmplx}.}
\end{figure}

Turning on more harmonics makes the model even more interesting. The decay of the amplitudes with $n$ is a well-documented phenomenon in critical, complex and fractal systems with the discrete scale invariance considered here \cite{GlSo}. Parametrizing the amplitudes as $A_n=a_n {\rme^{-\g n}}/{n^u}$ and $B_n=b_n {\rme^{-\g n}}/{n^u}$, one can plot the primordial scalar spectrum for $n_{\rm max}=O(10)$ harmonics. In this case, there may be constructive interference amplifying oscillations to log-periodic spikes deforming the classical general-relativistic spectrum (Fig.\ \ref{fig1}) \cite{cmplx}. The sensitivity to sharp features of present-day CMB data can constrain the $n$-dependence of the amplitudes efficiently and, from that, unravel important information on the microscopic structure of spacetime.

\paragraph*{Acknowledgments.} The author is under a Ram\'on y Cajal contract and is supported by the I+D grant FIS2014-54800-C2-2-P.

\end{document}